\documentclass{article}

\usepackage{PRIMEarxiv}

\usepackage[utf8]{inputenc} 
\usepackage[T1]{fontenc}    
\usepackage{hyperref}       
\usepackage{url}            
\usepackage{booktabs}       
\usepackage{amsfonts}       
\usepackage{nicefrac}       
\usepackage{microtype}      
\usepackage{lipsum}
\usepackage{fancyhdr}       
\usepackage{graphicx}       
\graphicspath{{media/}}     

\usepackage[polish, english]{babel}

\title{Complexity of Popularity and Dynamics of Within-Game Achievements in Computer Games}

\author{
  Leonardo R. Cunha*, Leonardo O. Mendes, and Renio S. Mendes \\
  Departamento de Física \\
  Universidade Estadual de Maringá \\
  Maringá, Paraná, Brazil \\
  \texttt{leoribeirocunha@gmail.com} \\
}

\begin{document}
\maketitle

\begin{abstract}
Tasks of different nature and difficulty levels are a part of people's lives. In this context, there is a scientific interest in the relationship between the difficulty of the task and the persistence need to accomplish it. Despite the generality of this problem, some tasks can be simulated in the form of games. In this way, we employ data from a large online platform, called Steam, to analyze games and the performance of their players. More specifically, we investigated persistence in completing tasks based on the proportion of players who accomplished game achievements. Overall, we present five major findings. First, the probability distribution for the number of achievements is log-normal distribution. Second, the distribution of game players also follows a log-normal. Third, most games require neither a very high degree of persistence nor a very low one. Fourth, players also prefer games that demand a certain intermediate persistence. Fifth, the proportion of players as a function of the number of achievements declines approximately exponentially. As both the log-normal and the exponential functions are memoryless, they are mathematical forms that describe random effects arising from the nature of the system. Therefore our first two findings describe random processes of fragmenting achievements and players while the last three provide a quantitative measure of the human preference in the pursuit of challenging, achievable, and justifiable tasks.
\end{abstract}


\section{Introduction}

We are continually faced with the most diverse tasks throughout our lives. Some of these tasks may be simple, complex or even impossible, and for each task, the individual's dedication to performing it depends on its degree of difficulty \cite{smith2015impact, chatelain2016task}. In this context, there is interest in understanding the relationship between difficulty, persistence and success \cite{dicerbo2014game, lucas2015going}. Persistence at a young age, for instance, has been shown to be predictive of academic outcomes, income and occupational level \cite{roberts2007power, poropat2009meta, andersson2011role}. Often, investigations into the success rate of completing tasks are not easy to carry out because the data about them are frequently difficult to obtain. In this work, instead of analyzing the success rate of tasks in general, we investigate this subject in the context of videogames, where we leverage a large available database.

Investigating aspects related to games is quite pertinent to today's society. They have continually gained in popularity and significance along recent years. Particularly, the games market is growing about $11\%$ per year with its revenues reaching almost $200$ billion dollars in present year, being substantially greater than the GDP of most countries \cite{newzoo2021}. In 2015, for instance, $60\%$ of USA inhabitants played digital games. The average player's age was about $35$ years old, the proportion of male and female players was $56\%$ and $44\%$, respectively \cite{entertainment2015essential}. In addition, videogames simulate many human tasks. As a consequence of games being constantly blended into our everyday activities, gamification is emerging in the most diverse contexts, such as work and physical activity training \cite{hamari2014measuring}.

This omnipresence of games in society has sparked scientific interest in multiple areas of knowledge \cite{qian2016game}. In particular, games are being employed in studies on human persistence. Consequently, theories have been implemented in an attempt to describe factors that affect individual motivation in games. Kahn et al. \cite{kahn2015trojan} developed and validated a scale intended for measuring video game-play motivation in general with six motivational components: socializer, completionist, competitor, escapist, story-driven, and smarty-pants. Other works adapt already consolidated approaches and apply them to the context of games. For example, in the field of psychology, the concept of flow refers to the ``process of optimal experience'' \cite{csikszentmihalyi1989optimal}. The emergence of flow is contingent upon critical boundary conditions: the individual’s level of skill in the respective domain; and the level of task challenge \cite{keller2008flow}. Based on this approach, for different games, it was demonstrated that different games affect player experience flow, and consequently their motivation \cite{jin2012toward}. This scientific literature presented makes clear the importance of better understanding aspects of human motivation. In a first instance, investigations can focus on games, however results obtained from such research can help to understand human motivation to perform more general tasks in life.

Despite many theories being proposed, the major challenge in conducting this type of research lies in obtaining reliable data. It is known that self-report data can be biased due to factors such as self-presentation motives, introspection inability, or the limitations of self-report tools \cite{hessing1988exploring, devaux2016social}. However, one way to replace these traditional methods of social science is through the use of automated methods of data collection, which enable entirely different scales of analysis \cite{newman2011structure, szell2010multirelational}. Nowadays, there are online platforms that sell and manage a large quantity of videogames: Epic Games \cite{Epic2020}, GOG \cite{GOG2020}, EA Origin \cite{Origin2020}, among others. One of the largest gaming platforms is Steam \cite{steam2020}, this platform has a representative sample of the player population, with millions of concurrent users distributed around the world. Steam has a catalog with more than $16$ thousand games. For the majority of these, developers insert progress markers throughout the game, called achievements. In this work based on Steam data, we consider the number of reached achievements (known as ``unlocked'' achievements in Steam) as a measure of success associated with player persistence (see more details in Methods).

Therefore, we studied the dynamics of completing tasks using Steam achievements. The novelty of the study is the use of a large database collected automatically, which makes it possible to simulate tasks. For each Steam game, we obtain the name, the total number of players, and the fraction of players that completed each achievement. All data correspond to games released from the start of the platform in 2004 to January 2021. In our work, we associate persistence as the main factor for unlocking achievements, because the feeling of motivation among players increases as they approach the achievement \cite{teng2017strengthening}. On the other hand, from a game perspective, we associate the proportion of players who unlocked achievements as a indicator associated with engaging factor. This nomenclature appears consistent since most online games are commercial systems designed to motivate gamers to continue playing by enticing them to progress from one game level to the next \cite{teng2018managing}. Thus, the main objective of the present work was to mathematically describe the proportion of players who unlocked game achievements.

We initially conducted an exploratory analysis of the global variables of the system, and fitted probability distributions to the number of achievements and the number of players per game. These two distributions are well characterized by a log-normal, which are common in other contexts of the natural and social sciences. During this exploratory study, we also found that game popularity (measured by the total number of players) is related to engaging factor. Next, we organized the games in our study in two distinct classes: the first concerning games that at least one player completed all achievements, and the second where no player completes all achievements. Using this grouping, we evaluate the fraction of players that completed all achievements (fraction of completists) and the maximum fraction of completed achievements by players (completed fraction). We focused separately on the shape of the probability distributions of the number of games as a function of the fraction of completists or in terms of the completed fraction. In particular, we identified games with more players are those that demand more persistence but not seemingly impossible. Advancing in our study, the fraction of players along the achievements of each game is investigated. Separating the games into groups with a similar engaging factor, an approximate exponential decay is identified over the game's achievements. Therefore, once again, we find a type of pattern that is common in studies of various natural and social systems \cite{butt1972study, norman1988tests, candia2019universal}.

\section{Methods}

The data employed in this work were collected from the Steam platform and are freely available. Steam is an online platform developed by Valve Corporation, which promotes the digital distribution and gives digital rights management of the products of its base \cite{steam2020}. Currently, Steam is one of the major platforms that offer such services, reaching more than twenty million online users simultaneously \cite{steam2019stats}. Furthermore, it records the activity of its users in an automated way. This process occurs without interference in the player activity. These facts indicate that the Steam data contain representative and reliable information about game activities.

For reasons of confidentiality and rights, the data available are particularly limited in order to protect users' identities. These data can be accessed via the Steam API \cite{steam2020dev} or from other sites that also provide some of these data. For instance, the site AStats \cite{astats2021} was used in this work to obtain the name of the games, number of players, and number of users who accomplished each achievement of the games considered. Data correspond to Steam players linked to AStats, thus, we do not have access to the achievement data of all players in Steam. The collected data refer to $16,806$ Steam games in February $2021$. Part of these data were not analyzed, due to missing information identified in the achievement records. After the data curation process, $15,239$ games remained to be investigated.

In games, an achievement is a meta-goal defined outside the parameters of the game. Achievements mark an accomplishment on the part of the player and are not necessarily linked to the main narrative of the game. Fulfilling the conditions of the achievement and receiving recognition of the achievement by the game is referred to as unlocking the achievement. It should be highlighted an exceptional case, some games offer the ``Ironman'' mode, wherein players face additional challenges or restrictions, often resulting in a more difficult and unforgiving experience. In games that offer this mode, it is possible to turn it on and off. If the ``Ironman'' mode is turned off, players can use cheats and features to make tasks within the game easier. However, only with the mode turned on is it possible to unlock the achievements. Unfortunately, the data we have access to do not distinguish this possibility and, therefore, our analysis does not incorporate this difference. One of the main goals of developers inserting achievements is to encourage players to keep playing beyond simply completing the main narrative of the game. In this way, the developer tries to increase the game's longevity by making the player find all its secrets and complete all its challenges. Thus, developers determine how many achievements there are in each game as well as what actions are necessary to accomplish each achievement. Curiously, a game with many achievements is not necessarily more difficult or time consuming than another with few achievements.

The interesting aspect of using achievements as a metric for gauging players' persistence resides in the visibility of these achievements within the player's online profile. The player's motivation for reaching achievements lies in maximizing their own overall score across games and gaining recognition for their performance due to publication of their achievement profiles. Some gamers pursue unlocking achievements as a goal in themselves, without particularly looking to enjoy the game, this community of players often refer to themselves as ``achievement hunters'' \cite{wilde2016life}. Taking advantage of this competitive scenario, we interpreted unlocking achievements as tasks and, in order to obtain quantitative results, we used the number of players as the main metric. We refer to the number of players as Steam users who have purchased, installed, and launched the game at least once. In addition, for each game in our database, we obtain the number of players that completed each achievement.

In this context, we consider achievements as success markers. Nonetheless, there can be several factors linked to people's decision to continue playing. Of all the factors, the literature focuses on three main ones: play with friends (social factor), intrinsic fun of the game (immersive factor), and challenge (individual factor) \cite{nick2006, Hoffman2010, boyle2012engagement, billieux2013you, dindar2018people, johnson2018validation}. Thus, for simplicity, we define that reaching achievement is the result of player persistence. Note that the definition of persistence used in this work summarizes all the factors responsible for keeping the player unlocking the achievements. From the game's point of view, the players' persistence set can be interpreted as an engaging factor. Once again, regardless of whether the game is fun or boring, whether it is easy or difficult, the game's engaging factor sums up all these factors and expresses the degree to which it keeps its players reaching for achievements. In general, we consider that a large proportion of players reaching the achievements in a given game signifies a high level of engagement among the players of the game, meaning the game exhibits a high engaging factor. Mathematically, we defined the engaging factor in two ways, because throughout the work we divided the games into two groups. The first group represents the games in which at least one player reached all available achievements, for this group we consider the engaging factor as the fraction of completists, that is, the fraction of players that unlocked all the achievements of the game. In the second group, we considered the games which no player completed all achievements, for this group the engaging factor is completed fraction, the fraction of achievements that the best player reached with success.

\section{Results}

\begin{figure*}[!ht]
\centering
\includegraphics[width=1\linewidth]{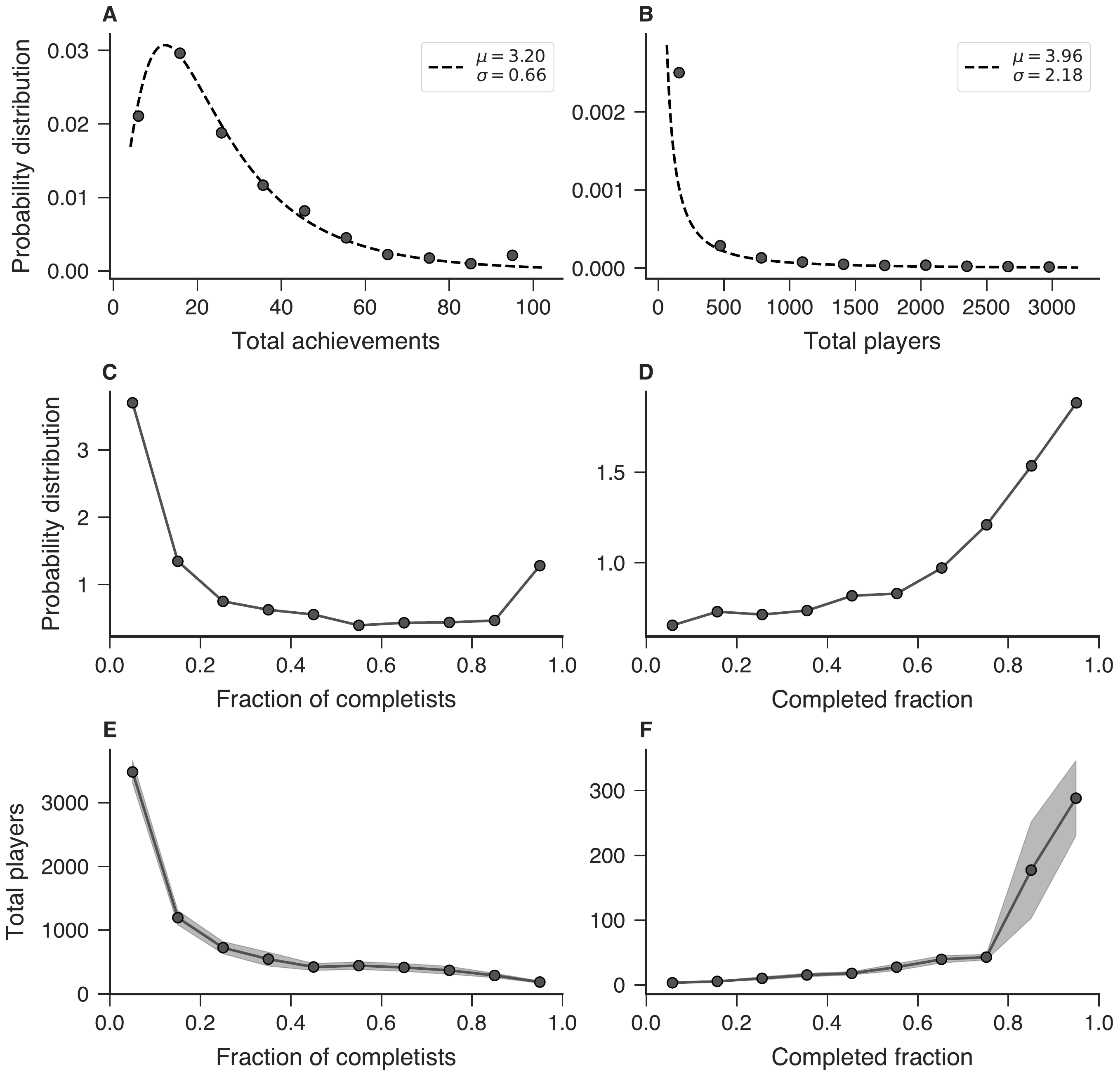}
\caption{{\bf Probability distribution of game and player variables.} Probability distributions of the number of games as function of the (A) number of achievements, (B) number of players, (C) fraction of completists, and (D) completed fraction. Panels (E) and (F) show the relationship between the fraction of completists and number of players and between the completed fraction and number of players, respectively. In panels (A), (B), (C), and (D) the markers represent the probability on the y-axis and the center of the bin on the x-axis. Dashed lines in (A) and (B) refer to fitted log-normal distributions, with parameters $\mu_A = 3.20$ [$95\%$CI: $3.18$, $3.22$] and $\sigma_A = 0.66$ [$95\%$CI: $0.64$, $0.67$], and $\mu_B = 3.96$ [$95\%$CI: $3.91$, $4.00$] and $\sigma_B = 2.18$ [$95\%$CI: $2.13$, $2.23$], respectively. In panels (E) and (F) the markers represent the average per window and the shaded area indicates the error relative to the average. In each case, the interval size was chosen to maintain $10$ markers.}
\label{fig:1}
\end{figure*}

We start by investigating the variables total achievements ($N$), total players ($G$), fraction of completists ($F$), and completed fraction ($U$). An inspection of the total achievements (the number of parts of a game) of each game shows that these totals vary from one to thousands. The maximum total achievements is $9,821$, corresponding to the game \textit{LOGistICAL}. The achievement distribution for all games is shown in Fig. \ref{fig:1}A. This distribution has a peak approximately at ten achievements, and there are $644$ games with this number of achievements. This figure also shows the fit of data via the log-normal distribution
\begin{equation}
    p(N)= \frac{1}{\sqrt{2 \pi} \, \sigma \, N} \, \exp{\left(-\frac{(\log N - \mu)^2}{2\sigma^2}\right)} ,
    \label{eq1}
\end{equation}
where $N$ stands for total achievements, $\log$ refers to the natural logarithm function, $\mu = 3.20$, and $\sigma = 0.66$. It is worth noticing that about $75$\% of data are encompassed within $40$ achievements, and $95$\% of the games have less than $100$ achievements. This fact is in line with the Steam recommendation that games should have a maximum of $100$ achievements \cite{hardwaresfera2020}.

A simple data inspection is enough to reveal that there is an abundance of games with a relatively small number of users and few games with many users. In a more detailed analysis of games with few players, we verify that the number of games decreases with the number of players. For instance, $647$ games have only one player, $397$ games with two, and $324$ games with three. In the other extreme, the most popular game, \textit{Counter-Strike: Global Offensive}, has $241,325$ players in our database. In addition, $78$\% of the games have less than $500$ users. Figure \ref{fig:1}B exhibits the distribution of number of games as a function of the number of players, where a clear decaying behavior is observed. This distribution is also in good agreement with a log-normal distribution, Eq. (\ref{eq1}), by substituting total achievements ($N$) for total players ($G$), with parameters $\mu = 3.96$ and $\sigma = 2.18$. As one can see, the tail of this log-normal is much longer than the one presented in Fig. \ref{fig:1}A, since the parameter $\sigma$ is less than in Fig. \ref{fig:1}B.

Due to variety of games, there is great diversity in the persistence degree required to unlock achievements. For example, the number of games in which at least one player has reached all achievements is $11,831$, comprising approximately $78$\% of the $15,239$ games in our database. In a first approximation, a possible metric to quantify the engaging factor needed to complete a given game is the fraction of users that completed it. As mentioned in Methods section, this fraction will be referred to as the fraction of completists, varying from $0$, for a game that no players completed, to $1$, for a game that all players complete. Considering only games that at least one player completed, Fig. \ref{fig:1}C shows the probability distribution of the number of games as a function of the fraction of completists. This distribution is concentrated at a small fraction of completists, that is, there are much more games only completed by a small fraction of players. Quantitatively, $6,076$ games correspond to the fraction of completists smaller or equal to $0.2$.

There are also $3,408$ games in which no player completed them, representing about $22$\% of our database. In this case, we quantify the engaging factor of a game as the fraction of achievements that the best player reached with success. Exactly as defined in the Methods section, this metric is referred to as the completed fraction. Thus, if the completed fraction is $0.6$ for a given game, only $60$\% of its achievements were unlocked considering all players. Among these games there are no completists, no game has a completed fraction equal to zero, $181$ present this fraction smaller than $0.1$, and $1,151$ have this fraction between $0.8$ and $1.0$. Figure \ref{fig:1}D displays the probability distribution of the number of games as a function of the completed fraction. For low values of completed fraction, the probability function increases slowly and monotonically. However, this ratio of increase becomes more intense for larger completed fraction.

Comparably to Figs. \ref{fig:1}C and \ref{fig:1}D, complementary information about our database is given in terms of total players as a function of engaging factor. These interrelated aspects are exhibited in Figs. \ref{fig:1}E and \ref{fig:1}F. As we can see, the patterns of Figs. \ref{fig:1}C and \ref{fig:1}D are essentially replicated in Figs. \ref{fig:1}E and \ref{fig:1}F, namely the popularity, estimated as number of games or total players, diminishes with the increase of the fraction of completists and grows with the increase of completed fraction. For instance, games with fractions of completists larger than zero and smaller or equal to $0.2$ accumulate $17,273,449$ players, corresponding to around $86$\% of all players. Moreover, games with the completed fraction larger or equal to $0.8$ collect $266,816$ users, which represent little more than $1$\% of all players.

\begin{figure*}[!ht]
\centering
\includegraphics[width=1\linewidth]{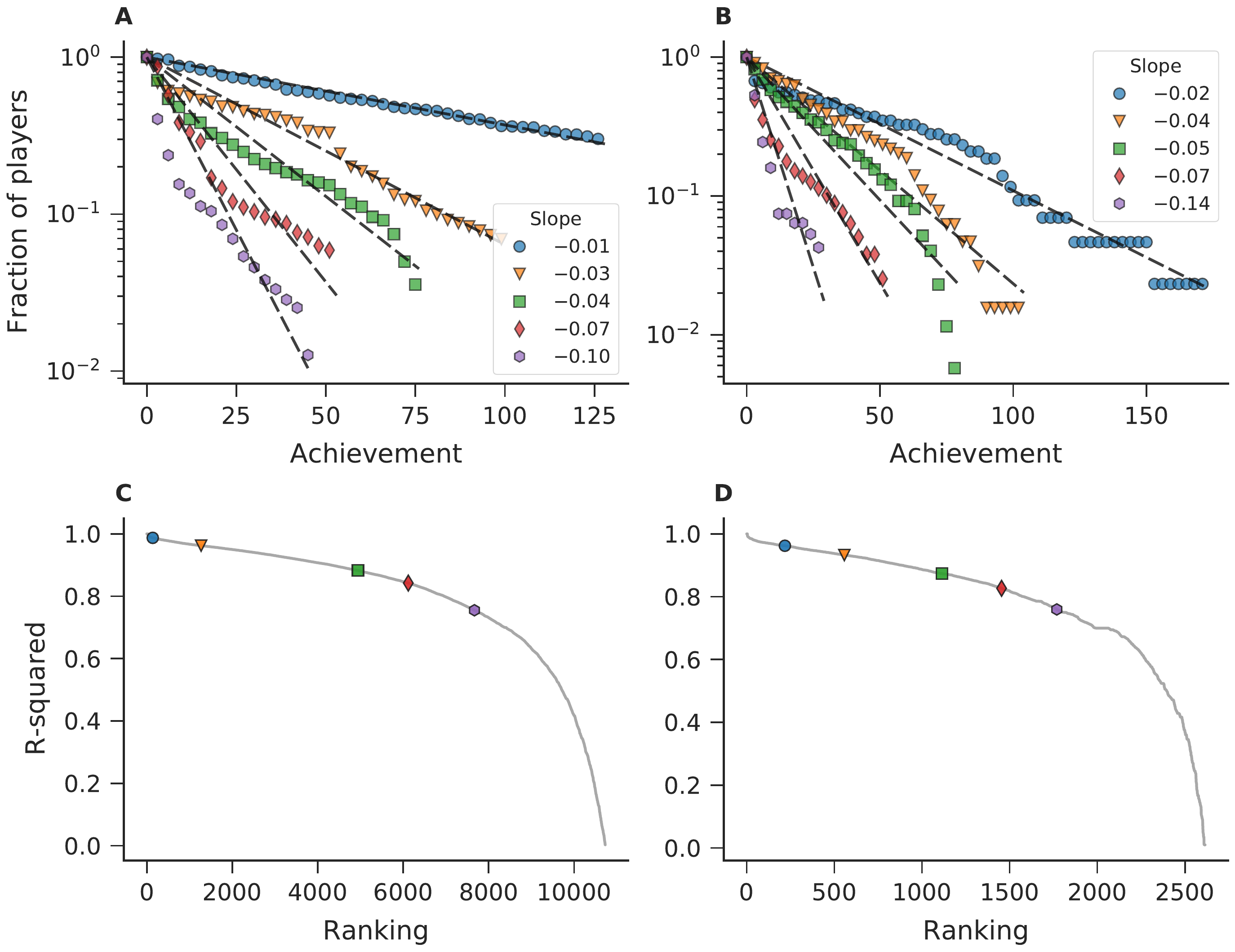}
\caption{{\bf Exponential decay of the fraction of players as a function of the unlocked achievements.} Examples of decay of the fraction of players as a function of the number of unlocked achievements for (A) completed games and (B) games where no user completed all achievements, in mono-log scale. In panel (B), we disregard achievements that no player has reached, given that we are working with the logarithmic scale. Markers refer to the data and straight lines are the corresponding exponential fits. R-squared coefficient for exponential fits according to ranking for (C) completed games and (D) games where no user completed all achievements. Colored markers correspond to R-squared values of curves shown in panels (A) and (B), and the light gray curve shows the behaviour for all games.}
\label{fig:2}
\end{figure*}

In the direction of advancing beyond overall engaging factor, we analyze the fraction of players for all game achievements. For each game, the achievements are organized in decreasing order of the number of players that unlocked them (for example, the first achievement was unlocked by a number of players, the second one by a smaller number, etc.). Mostly, the fraction of players is close to $1$ for the first achievement, indicating that almost all players reach a minimal success. An exploratory study of several games randomly chosen suggested exponential decays of this fraction of players as a function of the number of achievements. Taking advantage of the fact that exponential behaviors are nicely presented as straight lines in mono-log graphs, some of these random choices are illustrated in Figs. \ref{fig:2}A and \ref{fig:2}B. In addition, these dashed lines were fixed starting at fractions equal to one when the number of achievements is zero; this most simple choice was adopted because all players of any game correspond to a fraction equal to one. For a game with total achievements $N$, we suppose the variable $n$ represents a given number of achievements ($0 \leq n \leq N$). Thus, there is only one parameter to indicate the exponential decay of the fraction of players ($f(n)$), the slope ($\beta$) of its dashed line
\begin{equation}
    f(n) = \exp\left( - \beta n \right).
    \label{eq-exp}
\end{equation}
As we can see in Figs. \ref{fig:2}A and \ref{fig:2}B, some behaviors are close to exponential and others reveal clear deviations. Moreover, Fig. \ref{fig:2}A is associated to games with some fraction of completists. In turn, Fig. \ref{fig:2}B exemplifies games without any completists.

To quantify the quality of these exponential fits, the coefficient of determination R-squared ($R^2$) was employed \cite{miles2005r}. This coefficient expresses the percentage of variance of the data that is explained by the linear model. A null $R^2$ indicates that the model is as good as a constant line which equals the mean, and if $R^2$ equals one, it represents a model that explains all variation in the data. In order to restrict the analysis to a linear model, the logarithm of the data was employed throughout our study. Figure \ref{fig:2}C shows the ranking of $R^2$ for which there is a non-null fractions of completists, while Fig. \ref{fig:2}D exhibits the ranking of $R^2$ for all the games without any completists. These figures present a slow decay for small ranks, indicating that the majority of games have higher $R^2$. In particular, $8,378$ games have $R^2$ larger than $0.7$, corresponding to approximately $71$\% of the games with non-null fraction of completists. Additionally, for games that no player was able to conclude all achievements, $1,996$ present $R^2$ larger than $0.7$, which amounts to approximately $59$\% of these games. Note that Fig. \ref{fig:2}D contains only $2,616$ of the $3,408$ games where no player concluded all achievements. This limitation of data occurred because negative values of $R^2$ were omitted. This negative $R^2$ indicates that the fit is bad and the model is not able to describe the data \cite{barten1987coefficient}. These results point out that games with players that completed all achievements are usually better fitted by an exponential decay than the remaining games.

Despite the variability of players as a function of the achievements, we work towards the identification of average behavior associated with these decays. However, since there are many games with different numbers of achievements, we employ a normalized metric $x=n/N$, defined as the considered achievement $n$ divided by the total number $N$ of them. Therefore, the normalized achievement $x$ for a game assumes values between $0$ and $1$. Advancing in reasoning, for each game, the fraction of players that reached the proportion $x$ of achievements will be referred to as $f(x)$; thus, $f(x) = g(n)/G$, where $g(n)$ is the number of users that completed the $n$-th achievement and $G$ is the total of players. If $x=0$, all players of a game are included and we have $f(0)=1$ and $g(0)=G$. In turn, $f(1) = F$ and $g(N)=GF$.

\begin{figure*}[!ht]
\centering
\includegraphics[width=1.0\linewidth]{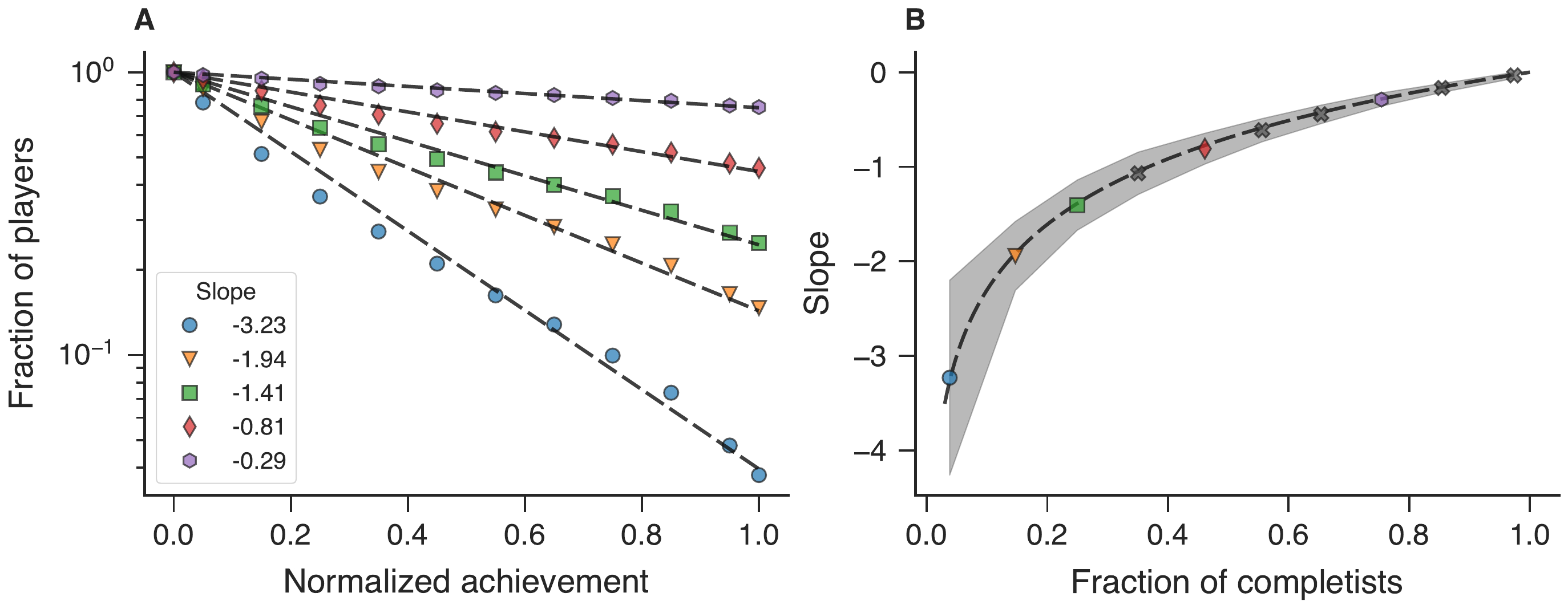}
\caption{{\bf Mean exponential decay for completed games.} (A) Examples of decay of the average fraction of players as function of normalized achievements in mono-log scale. The games were grouped in windows containing varying fractions of completists, in increments of $0.1$. Within each group, a mean series was constructed with $10$ equally spaced markers. Additionally, the first point of the mean series was fixed as $(0,1)$ and the last point as $(1,\mu_{F})$, where $\mu_{F}$ is the mean of the fractions of completists in the group. At the end of this procedure, the mean series of all groups had $12$ markers. Blue markers represent the average of the group of games with fraction of completists between $0$ and $0.1$, the orange markers greater between $0.1$ and $0.2$, the green markers between $0.2$ and $0.3$ , red markers between $0.4$ and $0.5$ and purple markers between $0.7$ and $0.8$. The dashed lines represent the exponential model fitted to each different group of completists. (B) Relationship between the exponential decay rate and the fraction of completists. In this way, we calculated the mean slope $\alpha$ and the mean of the fraction of completists $F$ for each group of games. The colors correspond to the groups shown in (A), that is, the blue maker is the slope of the group of games with a fraction of completists between $0.0$ and $0.1$, the orange marker is the slope of the group of games with fraction of completists between $0.1 $ and $0.2$, and so on. Gray markers represent games grouped with fractions of completists not exemplified in (A), but calculated in the same way. The gray shaded area shows a standard deviation band and the dashed line is the logarithmic function $\alpha =-\log(F)$.}
\label{fig:3}
\end{figure*}

To identify a collective behavior, we divided these data in windows and calculated the mean value of $f(x)$ for each window. That is, we grouped the games based on fractions of completists, in equally spaced and non-overlapping intervals of $0.1$. Within each of these groups, we calculated the mean series and fitted an exponential decay. Examples of this procedure is displayed in Fig. \ref{fig:3}A, where the colored markers correspond to the mean series and the dashed lines to the exponential fitted in each group. Because the figure is in mono-log scale, a straight line corresponds to an exponential decay. Taking into consideration that the mean series decays exponentially, we can write $f(x)=\exp(- \alpha x)$, where $\alpha$ refers to the slope of the line. Employing $f(1) = F$, one has $F=\exp(-\alpha )$ and $\alpha =-\log(F)$. In Fig. \ref{fig:3}B we depicted the slope $\alpha$ as a function of the fraction of completists $F$. The highlighted symbols represent the mean fraction of completists in each group and its respective slope ($\alpha$), the shade in gray marks one standard deviation up and down, and the dashed line is $-\log(F)$. Our results remain robust if the chosen interval size of the fractions of completists to group the games varies between $0.05$ and $0.25$. By using $x=n/N$ and $f(x)=g(n)/G$, we can alternatively write the empirical result $f(x)= \exp(-|\log(F)|x)$ as
\begin{equation}
    g_{c}(n) = G \, \exp\left( - \frac{|\log F|}{N} \, n \right),
    \label{eq2}
\end{equation}
where the index $c$ indicates completed games by players. Therefore, for each game, this result can be viewed as the mean number of players that reached the $n$-th achievement when there is a fraction of completists $F$, $N$ total achievements, and $G$ total players.

\begin{figure*}[!ht]
\centering
\includegraphics[width=0.5\linewidth]{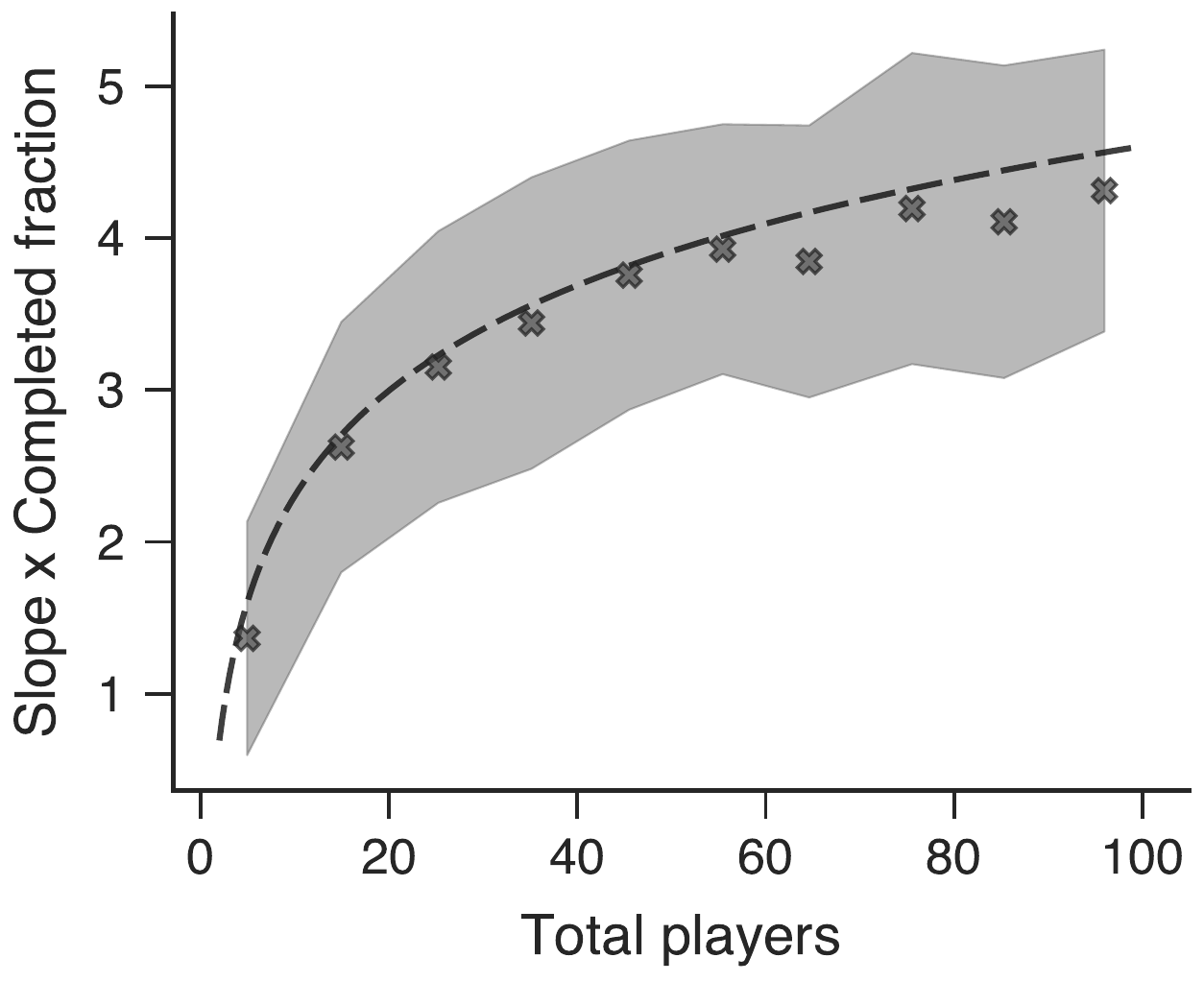}
\caption{{\bf Relationship between the exponential decay rate multiplied by completed fraction and total players.} For each game, the exponential decay rate $\alpha$ was calculated using the fraction of players on the y-axis and normalized achievement on the x-axis (all null values in the series were removed). Next, the ordered pair $(G, \alpha U)$ was taken, and the games were grouped into windows containing varying total players in increments of $10$. Within each window, the mean on the x and y axes was calculated. The gray markers represent the ordered pair of these means for each group. The gray shaded area shows standard deviation, and the dashed line represents the logarithmic function ($\alpha U = \log(G)$).}
\label{fig:4}
\end{figure*}

A similar line of reasoning as the one applied above can also be employed to investigate the mean behaviour of games without completists. As in the previous case, the number of players that started a game is $G$, that is, $g(0)=G$. For the vast majority of games without completists, we verified that only a few players $C$ of each game reached the maximum number of completed achievements, $n_{max}$. For instance, among the games without completists, there are approximately $82.2\%$, $9.4\%$ and $3.0\%$ of games with $C$ equal to one, two, and three, respectively. Thus, we exploit $g(n_{max}) = C$. Similarly and consistently with the exponential behaviors presented in Fig. \ref{fig:2}B, we consider $g(n)=G \exp(- \beta n)$, where $\beta$ is a parameter to be fitted. Since $g(0)=G$ and $g(n_{max}) = C$, one has $\alpha = \log(C/G)/n_{max}$. These results lead to
\begin{equation}
    g_{u}(n)=G \, \exp\left( - \frac{\log (G/C)}{n_{max}} \, n \right),
    \label{eq3}
\end{equation}
where the index $u$ stands for uncompleted games by players. Thus, we employ $g_u(n)$ as the mean numbers of players that reached to the $n$-th achievement for a game with $G$ gamers, and $C$ players that concluded the maximum number of achievements $n_{max}$. In this direction, the fraction of players can be obtained by defining $f(n) = g_{u}(n)/G$, and analogously to what was done in the previous case, $f(x) = \exp(-\log(G/C)x/U)$, where $x = n/N$ and $U = n_{max}/N$. Remembering that $f(x) = \exp(- \alpha x)$, we can write the mathematical relation $\alpha U = \log(G/C)$, where $\alpha$ is the slope of decay of the fraction of players along the normalized achievement. Figure \ref{fig:4} shows the average behavior of the relationship between these quantities. Since for $82.2\%$ of the games without completists, there is only one player that reached $n_{max}$ achievements, we can approximate $C$ by one ($C \approx 1$). Consequently, we approximate this relationship by the logarithmic model $\alpha U = \log(G)$ (dashed line).

\section{Discussion}

As previously mentioned, achievements are progress markers of games that as consequence promote their fragmentation in several parts. This fragmentation can present a certain degree of randomness, for instance, due to the fact that games are different from one another and the developers are also usually distinct. In these cases, we can think of starting from the whole and then doing successive partitions at random. Suppose a partition of size $p_1$ that is divided and one of its resulting components has a size $p_1 p_2$, where $p_2$ is the fraction of partitioning. For $n$ consecutive realizations of the fragmentation process, this result is extended to the product $p_1 p_2 \cdots p_n$, leading to a random multiplicative process. In situations such as this one, it is common to employ the log-normal, Eq. (\ref{eq1}), to describe the distribution of the sizes \cite{crow1987lognormal}. Particularly, there are social systems that can be divided into parts with some degree of randomness and modeled by log-normal distributions such as, for instance, the number of votes received by candidates \cite{fortunato2007scaling}, popularity dynamics of memes \cite{bauckhage2011insights}, and digg values \cite{van2011lognormal}. Examples like these and the theoretical approach above motivated us to model our data for the distribution of games as a function of the total achievements by a log-normal, shown in Fig. \ref{fig:1}A. Alternatively, we have attempted to fit other distributions, such as Weibull, gamma, and Burr. Nonetheless, with the log-normal, we obtained the best Kolmogorov-Smirnov statistic.

As players are distributed across the different games, we can consider a model of partitioning of these users. In this direction, the previous model can be viewed as a first approximation. However, we consider that all players are partitioned across the games instead of each game partitioned in achievements as in the above paragraph. This reasoning inspired us again to employ the log-normal, Eq. (\ref{eq1}), to model the distribution of games as a function of the total players, as presented in Fig. \ref{fig:1}B. As it was remarked in connection to Fig. \ref{fig:1}A, other distributions were tested, but the log-normal had the best Kolmogorov-Smirnov statistic.

Taking advantage of fraction of completists and completed fraction definitions, identified characteristics of relating to most games and to the most popular ones. For games fully completed by some players, the most common games are those that have lower engaging factors (Fig. \ref{fig:1}C). Considering games not completed by any player, the most frequent ones are those closest to completion (Fig. \ref{fig:1}D). Concerning popularity, Figs. \ref{fig:1}E and \ref{fig:1}F indicate similar patterns considering players instead games. These results obtained seem consistent with previous research. If we consider the engaging factor as a measure of the average degree of persistence (associated to difficulty) to complete all the achievements, then our work reinforces the idea that motivation grows with increasing task difficulty, while success is possible and justified \cite{lucas2015going, chatelain2016task, duchowski2018index}. Strojny and colleagues found that people's involvement increases as the game becomes more difficult, but after a certain ``optimal'' difficulty involvement drops off quickly \cite{strojny2023player}. Furthermore, it seems clear the developers' intention is to define achievements aiming for this ``optimal'' degree of persistence. Possibly, the developers do this to motivate players to keep playing and consuming everything the game has to offer.

Before discussing the results related to Figs. \ref{fig:2}, \ref{fig:3}, and \ref{fig:4}, we consider a framework for discussing persistence in performing the different steps of a task. Particularly, we can think about the relative engaging of two different steps. This relative engaging can be defined as the number of people that finalized a step divided by the remaining number. Note that in this comparison, it is advantageous to consider situations with many people executing a given task in order to overcome undesirable statistical fluctuations. If the steps of a task are organized in decreasing order as achievements are systematized in this work, it is natural to consider the relative engaging of successive steps. We employ $r_i$ to denote the relative engaging of the $i$-th step in comparison to the previous one. Thus, $r_i=n_i / n_{i-1}$ with $i=1,2, \cdots$, where $n_i$ refers to the number of people that finalized the $i$-th step and, particularly, $n_0$ is the number of people that started the task. Consequently, $r_i$ assumes values from zero to one, zero (one) when the number of people who conclude the $i$-th task is null (did not change). In general, we have $n_1= n_0 \, r_1$, $n_2= n_1 \, r_2$ and so on. As a consequence, the number of people that ends the $i$-th step can be written in terms of the relative engagings and initial number of people that started the task, namely $n_i= n_0 \, r_1 \, r_2 \, \cdots \, r_i$. Applied to games, the relative engaging of the last achievement of a game and of starting the game coincides with the fraction of completists. By using the above notation, this fraction is equal to the ratio $n_N/n_0$ and, consequently, reduces to the product $r_1 \, r_2 \cdots r_N$. Since the fraction of completists depends on several factors, this last relationship indicates that the relative engagings are also connected to them. Moreover, the values of the successive relative engagings may oscillate significantly and exhibit a complex profile. Notably, the simplest pattern of these relative difficulties occur when they are equal, that is, $r_i=r$ for all $i$’s with $r$ constant. For this simplest model, one has $n_i=n_0 \, r^i$ or, alternatively, the exponential decay $n_i=n_0 \exp(-b \, i)$ with $b$ being a positive constant equals to $-\log(r)$, since $r$ assumes values between $0$ and $1$.

Returning to our results, they indicated that the exponential model exposed above is a good approximation for the majority of games (see Fig. \ref{fig:2}). Interestingly, this exponential model can be qualified as memoryless \cite{csiszar2011information}. In fact, if the relative engaging is independent of the pair of consecutive achievements considered in a model ($r_i=r$), we cannot identify (remember) a pair from the $r_i$'s and, consequently, the model can be referred as memoryless. In this direction, more deviations from the exponential model indicate a distancing from memorylessness, that is, more memory effects are expected. Consistently with this framework, one could define a memoryless index as the coefficient of determination for the straight line fit as in Figs. \ref{fig:2}A and \ref{fig:2}B. Thus, the maximum value for this index is equal to one for a completely memoryless model. From this definition also follows that, this index is only correlated by deviations from the exponential (that is, coefficient of determination R-squared), it is not susceptible to its rate decay (value of $r$).

We verified that, even after grouping games, the exponential model also provides a valuable description, as evidenced in Figs. \ref{fig:3} and \ref{fig:4}. Particularly, in the case of incomplete games, there is a deviation for large number of players as can be viewed in fig. \ref{fig:4}. Including, we present the average behavior of games with up to $100$ total players, which accounts for $89\%$ of games without completists, because for the remaining games, the fluctuation becomes too significant. Observing the fig. \ref{fig:4}, the model overestimates the real data (the dashed line is above the mean value obtained from the data). This occurs because the dashed line represent logarithmic function with $C=1$ (the most frequent case), but $C$ can assume values greater than one, since it is proportional to the total players in a game. Finally, the average exponential behavior in Eqs. (\ref{eq2}) and (\ref{eq3}) can be viewed as players dropping out at a fixed rate, where this rate depends on the engaging factor of the game.

Despite the large dataset and statistically robust results, our work also has some limitations. Above all, the data used does not distinguish specific conditions of games and players. As previously mentioned, some games feature the Ironman mode, and our data does not differentiate such games. Possibly for these games, players often choose to keep the Ironman mode turned off. In this condition, players can still spend time and persist in pursuing objectives within the game; however, achievements will not be unlocked. Given this, we believe that games like these may experience more faster exponential decay over achievements, as players have the option to turn off achievement unlocking. Alternatively, these achievements from a portion of players that are not counted — if they were counted — could contribute to the emergence of other mathematical patterns. Similarly, slightly different patterns may arise simply due to an intrinsic difference in the game, considering that games have different genres, styles, and qualities. In addition to game characteristics, our data does not differentiate the players, ranging from casual players to the aforementioned ``achievement hunters''. We still believe that using such a general dataset, as employed in this work, provides a perspective on the global pattern of human persistence. However, if the purposes and motivations of the players were known, it could lead to more practical and targeted results. Digging even deeper into this discussion, the factors that drive a person's motivation to continue playing can be dynamic over time \cite{schultheiss2007long, reinhard2009situational}. That is, a large-scale database that provides data on players' satisfaction over time could be highly beneficial from a scientific standpoint. Although we do not consider new games being published daily in this work, users may simply lose interest and transition to newer experiences not because of a perceived lack of challenge or boredom, but due to the novelty factor of other releases.

In the current work, we focused on determining mathematical patterns in the dynamics of completing tasks in Steam games. Although the used definitions of ``persistence'' and ``engaging factor'' summarize many factors, we believe that the results obtained in this work contribute to the understanding of the dynamics of human persistence in performing tasks. Future research can establish connections between the engaging factor and scales for measuring human persistence, such as those mentioned in the Introduction. Or even other scales based on the time spent and the number of attempts to solve a logic task \cite{shute2013measuring, ventura2013relationship}. Connecting theory to large-scale data can provide a improvement in the development of teaching and learning techniques. In the field of pedagogy, for instance, quantitative results about the dynamics of completing tasks could be implemented to maximize people's learning. That is, based on a person's performance and persistence, the difficulty of tasks could be regulated to maximize learning effects. Focus to economic interests, for the games industry, understanding the correlation between the engaging factor and the success of a game could be helpful. Furthermore, quantitative results help to better comprehend the average consumption behavior of players, which can be useful for improving sales strategies and thus maximizing profits. In general, the database used in the current work is a representative sample of players from all over the world, but future work may test the validity of our findings using even other databases.

\section{Conclusion}

In this work, we presented results about videogames and their players. We take advantage of Steam's large database, with over $16$ thousand games, and data automatically logged without the need for self-report to arrive at robust conclusions. Our results showed general quantitative characteristics of the structuring of the game achievements and contribute to the debate on the popularity of cultural objects by discussing the popularity of games. Finally, analyzing the proportion of players that obtain each achievement allows us to analyze the engaging factor of the games and player persistence. Summarizing the five conclusions of the current work: the probability distribution of the number of game achievements is well described by a log-normal; the distribution of game popularity also follows approximately a log-normal; developers prefer to make games that require neither too much nor too little perseverance on the part of the player; players also prefer games that demand a certain intermediate persistence; the number of players as a function of completed achievements decays approximately exponentially. Particularly, we believe that our findings can be useful for the gaming industry, for further understanding the relationship between game challenge and player persistence. In addition, investigations into the general characteristics of games lead to a better understanding of the average consumption behavior of players. The database used in the current work is a representative sample of players from all over the world, but future work may test the validity of our findings using even larger and more complete databases.

\section*{Acknowledgments}

This research was supported by Coordination for the Improvement of Higher Education Personnel (CAPES) and National Council for Scientific and Technological Development (CNPq -- Grants 407690/2018-2 and 303121/2018-1).

\section*{Author contributions}

All authors have contributed equally to the work reported on in this paper.

\textbf{Data Availability Statement} A data is available upon request from the corresponding author.


\end{document}